\documentclass[12pt]{iopart}
\usepackage{iopams}
\usepackage{graphicx}
\usepackage{cite}

\begin{document}

\title[Coupled nonlinear stochastic differential equations generating $1/f$
noise]{Coupled nonlinear stochastic differential equations generating arbitrary
distributed observable with $1/f$ noise}

\author{J Ruseckas, R Kazakevi\v{c}ius and B Kaulakys}

\address{Institute of Theoretical Physics and Astronomy, Vilnius University,
A.~Go\v{s}tauto 12, LT-01108 Vilnius, Lithuania}

\ead{julius.ruseckas@tfai.vu.lt}

\begin{abstract}
Nonlinear stochastic differential equations provide one of the mathematical
models yielding $1/f$ noise. However, the drawback of a single equation as a
source of $1/f$ noise is the necessity of power-law steady-state probability
density of the signal. In this paper we generalize this model and propose a
system of two coupled nonlinear stochastic differential equations. The equations
are derived from the scaling properties necessary for the achievement of
$1/f^\beta$ noise. The first equation describes the changes of the signal,
whereas the second equation represents a fluctuating rate of change. The
proposed coupled stochastic differential equations allows us to obtain
$1/f^\beta$ spectrum in a wide range of frequencies together with the almost
arbitrary steady-state density of the signal. 

\end{abstract}

\submitto{{\it J. Stat. Mech.}}

\maketitle

\section{Introduction}

Noise plays an essential role in many physical, biological and even social
systems. Therefore, for the understanding of those systems it is important to
characterize the noise and explain its origin. One of the characteristics used
for description of the noise is the power spectral density (PSD). In may cases
the noise can be modeled as a white noise which has a frequency-independent PSD.
However, there are various physical systems where the noise has significant
dependence on frequency. The characteristic behavior of the PSD is referred to
as a ``color'' of the noise. Pink noise or $1/f$ noise is a random process
described by the PSD $S(f)$ inversely proportional to the frequency, that is
$S(f)\propto1/f^{\beta}$ with $\beta$ close to $1$. $1/f$ noise has been
observed first as an excess low-frequency noise in vacuum tubes
\cite{Johnson1925,Schottky1926}. Later such noise has been found in condensed
matter
\cite{Bernamont1934,Bernamont1937a,Bernamont1937b,McWhorter1957,Hooge1981} and
other systems \cite{Weissman1988,Mandelbrot-1999,Scholarpedia2007}. Origin and
the general nature of $1/f$ noise is up to now the subject of discussions and
investigations, for review see
\cite{Wong2003,Scholarpedia2007,Kogan2008,Balandin2013}.

Many models have been proposed to explain the origin of $1/f$ noise; for a short
overview of the models see introduction of \cite{Kaulakys2009}. In many
condensed matter systems the $1/f$ spectrum is considered as a superposition
Lorentzians with a wide range distribution of relaxation times
\cite{Bernamont1937b,McWhorter1957,Ralls1984,Rogers1984,Kaulakys2005,Watanabe2005}.
In this approach $1/f^{\beta}$ noise with the desirable slope $\beta$ requires a
certain distribution of parameters of the system
\cite{Hooge1981,Dutta1981,Weissman1988,VanVliet1991,Wong2003,Kaulakys2005}.
However, it has been shown that only several well separated decay rates are
sufficient to yield an approximately $1/f$ power spectrum \cite{Erland2011}.
Self-organized criticality (SOC) provides models of $1/f$ noise relevant for
understanding of driven non-equilibrium systems \cite{Bak1987,Bak1996}. The
mechanism of SOC not necessarily yields $1/f$ fluctuations
\cite{Jensen1989,Kertesz1990}. The $1/f$ noise in the fluctuations of a mass was
first seen in a sandpile model with threshold dissipation proposed in
\cite{Ali1995} and was analytically obtained in a one-dimensional directed model
of sandpiles \cite{Maslov1999}. Yet another model of $1/f$ noise represents the
signals as sequences of the renewal pulses or events with the power-law
distribution of the inter-event time \cite{Lowen2005}. Recently, thermal
finite-size fluctuations as mechanism for $1/f$ noise has been proposed
\cite{Chamberlin2014}.

In some systems the $1/f$ fluctuations are non-Gaussian
\cite{Orlyanchik2008,Melkonyan2010}. Power-law distribution of signal intensity
as well as power-law behavior of the PSD in a wide range of frequencies can be
obtained using point processes where the time between the adjacent pulses
experience relatively slow the Brownian-like motion
\cite{Kaulakys1998,Kaulakys1999,Kaulakys2000-2}. Starting from this point
process model nonlinear stochastic differential equations (SDEs) generating
$1/f^{\beta}$ noise have been derived in
\cite{Kaulakys2004,Kaulakys2006,Kaulakys2009}. Such nonlinear SDEs have been
applied to describe signals in socio-economical systems
\cite{Gontis2010,Mathiesen2013} and as a model of neuronal firing \cite{Ton2015}.

However, in most cases $1/f$ noise is a Gaussian process
\cite{Kogan2008,Li2012}. The drawback of the nonlinear SDEs generating signals
with $1/f^{\beta}$ PSD, proposed in \cite{Kaulakys2004,Kaulakys2006}, is the
necessity of power-law steady-state probability density function (PDF) of the
signal. It is impossible to obtain Gaussian PDF together with $1/f$ spectrum
from such nonlinear SDEs. The purpose of this paper is to remedy this drawback
of nonlinear SDEs as source of $1/f$ noise by considering not only one SDE, but
a system of two coupled SDEs. In the system of coupled SDEs we interpret the
first equation as giving the signal, whereas the second equation represents a
fluctuating rate of change. We demonstrate that the proposed coupled stochastic
differential equations allows us to obtain $1/f$ spectrum in a wide range of
frequencies together with almost arbitrary steady-state PDF of the signal.

The paper is organized as follows: in \sref{sec:derivation}
we obtain a system of coupled SDEs generating signals with $1/f^{\beta}$
PSD by considering the scaling properties of the equations. Numerical
methods of solution of such equations are discussed in \sref{sec:numer}.
SDEs obtained in \sref{sec:derivation} do not have the most
general form that is allowed by scaling properties required to get
$1/f^{\beta}$ spectrum. For completeness, in \sref{sec:general}
we consider a more general, but more complicated form of equations.
\Sref{sec:concl} summarizes our findings.

\section{Derivation of coupled stochastic differential equations using scaling
properties}

\label{sec:derivation}In this section we obtain a pair of coupled nonlinear SDEs
by considering the scaling properties required to get $1/f^{\beta}$ PSD. The
method we use is similar to that in \cite{Ruseckas2014}, however now we
consider two stochastic variables and two equations. We assume that the first
equation describes the fluctuations of the signal, with the fluctuating rate of
change described by the second equation.

We can obtain a pair of coupled nonlinear SDEs generating signals exhibiting
$1/f$ noise by using the following considerations. The Wiener-Khintchine theorem
\begin{equation}
C(t)=\int_{0}^{+\infty}S(f)\cos(2\pi f)\rmd f\label{eq:wiener-khintchine}
\end{equation}
relates the PSD $S(f)$ to the autocorrelation function $C(t)$. If the PSD has a
power-law behavior $S(f)\sim f^{-\beta}$ in a wide range of frequencies
$f_{\mathrm{min}}\ll f\ll f_{\mathrm{max}}$, then, when the influence of the
limiting frequencies $f_{\mathrm{min}}$ and $f_{\mathrm{max}}$ is neglected, the
PSD has a scaling property
\begin{equation}
S(af)\sim a^{-\beta}S(f)\label{eq:spectr-scaling}
\end{equation}
for the frequencies in this range. In this paper we will consider signals with
PSD having $1/f^{\beta}$ behavior only in some wide intermediate region of
frequencies $f_{\mathrm{min}}\ll f\ll f_{\mathrm{max}}$. To avoid the divergence
of the total power occuring for pure $1/f$ behavior at arbitrarily small
frequencies we assume that the PSD is bounded for small frequencies $f\ll
f_{\mathrm{min}}$ outside of this region. Compatibility with experimental data
can be ensured by choosing sufficiently small limiting frequency
$f_{\mathrm{min}}$.

From the Wiener-Khintchine theorem \eref{eq:wiener-khintchine} and
equation \eref{eq:spectr-scaling} it follows that the autocorrelation function has
the scaling property
\begin{equation}
C(at)\sim a^{\beta-1}C(t)\label{eq:autocorr-scaling}
\end{equation}
in the time range $1/f_{\mathrm{max}}\ll t\ll1/f_{\mathrm{min}}$. Assuming that
we have two stochastic variables $x$ and $y$ with the signal represented by the
stochastic variable $x$, the autocorrelation function can be written as
\cite{Ruseckas2010,Risken1996,Gardiner2004}
\begin{equation}
\fl C(t)=\int \rmd x\rmd y\int \rmd x'\rmd y'\,xx'P_{0}(x,y)P(x',y',t|x,y,0)
-\left[\int \rmd x\rmd y\,xP_{0}(x,y)\right]^{2}\,.\label{eq:autocorr}
\end{equation}
Here $P_{0}(x,y)$ is the steady-state PDF and $P(x',y',t|x,y,0)$ is the
transition probability (the conditional probability that at time $t$ the
stochastic variables have values $x'$ and $y'$ with the condition that at time
$t=0$ they had had the values $x$ and $y$). The transition probability can be
obtained from the solution of the Fokker-Planck equation with the
initial condition $P(x',y',0|x,y,0)=\delta(x-x')\delta(y-y')$. The last term in
equation \eref{eq:autocorr}, being a constant, does not influence the PSD at
frequencies $f>0$. Therefore, we will neglect this term from now on.

One of the ways to obtain the required scaling property
\eref{eq:autocorr-scaling} is for the steady-state PDF to be a power-law
function of the stochastic variable $y$,
\begin{equation}
P_{0}(x,y)\sim p(x)y^{-\lambda}\,,\label{eq:steady-pdf}
\end{equation}
and for the transition probability to have the scaling property
\begin{equation}
aP(x',ay,t|x,ay,0)=P(x',y',a^{\mu}t|,x,y,0)\,.\label{eq:transition-scaling}
\end{equation}
Here $\mu$ is the scaling exponent and $\lambda$ is the power-law exponent of
the steady-state PDF of the stochastic variable $y$.
\Eref{eq:transition-scaling} means that the change of the magnitude of the
stochastic variable $y\rightarrow ay$ is equivalent to the change of time scale
$t\rightarrow a^{\mu}t$. Using
equations \eref{eq:autocorr}--\eref{eq:transition-scaling} and performing a change
of variables we get
\begin{eqnarray}
C(at) & = \int \rmd x\rmd y\int \rmd x'\rmd y'xx'P_{0}(x,y)P(x',y',at|x,y,0)\\
& \sim \int \rmd x\rmd y\int \rmd x'\rmd y'xx'p(x)y^{-\lambda}a^{\frac{1}{\mu}}
P(x',a^{\frac{1}{\mu}}y',t|x,a^{\frac{1}{\mu}}y,0)\\
& \sim a^{\frac{\lambda-1}{\mu}}\int \rmd x\rmd u\int \rmd x'\rmd u'xx'p(x)u^{-\lambda}
P(x',u',t|x,u,0)\,.
\end{eqnarray}
Therefore, the autocorrelation function has the required scaling property
\eref{eq:autocorr-scaling} with $\beta$ given by
\begin{equation}
\beta=1+\frac{\lambda-1}{\mu}\,.\label{eq:beta-1}
\end{equation}
We see that we obtain the pure $1/f$ noise when $\lambda=1$.

In order to avoid the divergence of the steady-state PDF \eref{eq:steady-pdf},
the diffusion of stochastic variable $y$ should be restricted at least from the
side of small values. In general, equation \eref{eq:steady-pdf} can hold only in
some region $y_{\mathrm{min}}\ll y\ll y_{\mathrm{max}}$. When the diffusion of
stochastic variable $y$ is restricted, equation \eref{eq:transition-scaling} also
cannot be exact. However, if the influence of the limiting values
$y_{\mathrm{min}}$ and $y_{\mathrm{max}}$ can be neglected for the time $t$ in
some region $t_{\mathrm{min}}\ll t\ll t_{\mathrm{max}}$, we can expect for the
scaling \eref{eq:autocorr-scaling} to be approximately valid in this time
region.

To get the required scaling \eref{eq:transition-scaling} of the
transition probability, only powers of the stochastic variable $y$
should enter into the pair of SDEs. Assuming that the coefficient in the
noise term of the first equation is proportional to $y^{\eta}$, we
will consider the following coupled It\^o SDEs 
\begin{eqnarray}
\rmd x_{t} = a(x_{t})y_{t}^{2\eta}\rmd t+b(x_{t})y^{\eta}\rmd W_{t}\,,
\label{eq:sde-x-1}\\
\rmd y_{t} = u(x_{t})y_{t}^{2\eta+1}\rmd t+\sigma y_{t}^{\eta+1}\rmd W_{t}^{\prime}\,.
\label{eq:sde-y-1}
\end{eqnarray}
Here $W_{t}$ and $W_{t}^{\prime}$ are standard Wiener processes.
The parameter $\sigma$ in equation \eref{eq:sde-y-1} gives the intensity
of the noise and the coefficient $u(x)$ needs to be determined. One
can see that equations \eref{eq:sde-x-1} and \eref{eq:sde-y-1} indeed
lead to the scaling of transition probability \eref{eq:transition-scaling}.
Changing the variable $y$ in \eref{eq:sde-x-1}, \eref{eq:sde-y-1}
to the scaled variable $y_{\mathrm{s}}=ay$ or introducing the scaled
time $t_{\mathrm{s}}=a^{2\eta}t$ and using the property of the Wiener
process $\rmd W_{t_{\mathrm{s}}}\stackrel{d}{=}a^{\eta}\rmd W_{t}$ we get
the same resulting equations. Therefore, the change of the scale of
the variable $y$ and change of time scale are equivalent, as in equation
\eref{eq:transition-scaling}, and the scaling exponent $\mu$ is
equal to
\begin{equation}
\mu=2\eta\,.
\end{equation}

To ensure steady-state PDF \eref{eq:steady-pdf} and for determination the
unknown coefficient $u(x)$ in equation \eref{eq:sde-y-1} we write the Fokker-Planck
equation corresponding to the system of SDEs \eref{eq:sde-x-1} and
\eref{eq:sde-y-1} \cite{Gardiner2004}
\begin{equation}
\fl\frac{\partial}{\partial t}P=-y^{2\eta}\frac{\partial}{\partial x}a(x)P
-u(x)\frac{\partial}{\partial y}y^{2\eta+1}P+\frac{1}{2}y^{2\eta}
\frac{\partial^{2}}{\partial x^{2}}b^{2}(x)P+\frac{1}{2}\sigma^{2}
\frac{\partial^{2}}{\partial y^{2}}y^{2\eta+2}P\,.
\end{equation}
The steady-state PDF $P_{0}(x,y)$ is the solution of the equation
\begin{equation}
\fl -y^{2\eta}\frac{\partial}{\partial x}a(x)P_{0}
-u(x)\frac{\partial}{\partial y}y^{2\eta+1}P_{0}+\frac{1}{2}y^{2\eta}
\frac{\partial^{2}}{\partial x^{2}}b^{2}(x)P_{0}+\frac{1}{2}\sigma^{2}
\frac{\partial^{2}}{\partial y^{2}}y^{2\eta+2}P_{0}=0\,.\label{eq:stationary-FP}
\end{equation}
\Eref{eq:stationary-FP} can be written in terms of the components
of the probability current
\begin{eqnarray}
J_{x}(x,y) = y^{2\eta}a(x)P_{0}-\frac{1}{2}y^{2\eta}
\frac{\partial}{\partial x}b^{2}(x)P_{0}\,,\label{eq:sx}\\
J_{y}(x,y) = u(x)y^{2\eta+1}P_{0}-\frac{1}{2}\sigma^{2}
\frac{\partial}{\partial y}y^{2\eta+2}P_{0}\label{eq:sy}
\end{eqnarray}
as
\begin{equation}
\frac{\partial}{\partial x}J_{x}(x,y)+\frac{\partial}{\partial y}J_{y}(x,y)=0\,.
\end{equation}
Inserting equation \eref{eq:steady-pdf} into \eref{eq:sx} and
\eref{eq:sy} we get
\begin{eqnarray}
J_{x}(x,y) = y^{2\eta-\lambda}\left[a(x)p(x)
-\frac{1}{2}\frac{\rmd}{\rmd x}b^{2}(x)p(x)\right]\,,
\label{eq:sx-1}\\
J_{y}(x,y) = y^{2\eta+1-\lambda}p(x)\left[u(x)-\sigma^{2}\left(\eta+1
-\frac{\lambda}{2}\right)\right]\,.
\label{eq:sy-1}
\end{eqnarray}
Assuming that the $x$-component of the probability current $J_{x}$
should vanish at the reflective boundaries that are not parallel to
$x$ axis, we get that the expression in the square brackets in equation \eref{eq:sx-1}
should be zero for different values of $y$. Thus the function $p(x)$
in \eref{eq:steady-pdf} should be a solution of the differential
equation
\begin{equation}
a(x)p(x)-\frac{1}{2}\frac{\rmd}{\rmd x}b^{2}(x)p(x)=0\,.
\end{equation}
This equation means that the steady-state PDF of the stochastic variable
$x$ is determined only by the coefficients $a(x)$ and $b(x)$ of
the SDE~\eref{eq:sde-x-1}. Further, assuming that the $y$-component
of the probability current $J_{y}$ should vanish at the boundaries
that are not parallel to $y$ axis, we get that the expression in
the square brackets in equation \eref{eq:sy-1} should be zero for different
values of $x$. Therefore, $u(x)=\sigma^{2}(\eta+1-\lambda / 2)$
and the required system of coupled SDEs is
\begin{eqnarray}
\rmd x_{t} = a(x_{t})y_{t}^{2\eta}\rmd t+b(x_{t})y_{t}^{\eta}\rmd W_{t}\,,
\label{eq:sde-x-2}\\
\rmd y_{t} = \sigma^{2}\left(\eta+1-\frac{\lambda}{2}\right)y_{t}^{2\eta+1}\rmd t
+\sigma y_{t}^{\eta+1}\rmd W_{t}^{\prime}\,.
\label{eq:sde-y-2}
\end{eqnarray}
Note, that the second equation \eref{eq:sde-y-2} has the form of
non-linear SDEs proposed in \cite{Kaulakys2004,Kaulakys2006}. Equations
similar to \eref{eq:sde-x-2}, \eref{eq:sde-y-2} have been considered
in \cite{Kaulakys2013}. From equation \eref{eq:beta-1} it follows that
the power-law exponent in the PSD of the signal generated by SDEs
\eref{eq:sde-x-2}, \eref{eq:sde-y-2} is related to the parameters
$\eta$ and $\lambda$ as 
\begin{equation}
\beta=1+\frac{\lambda-1}{2\eta}\,.\label{eq:beta-2}
\end{equation}

To get a stationary process and avoid the divergence of steady-state PDF,
equation \eref{eq:sde-y-2} should be considered together with boundaries restricting
the diffusion of stochastic variable $y$ or be modified. The simplest choice
restricting the range of diffusion of the stochastic variable $y$ is the
reflective boundaries at $y=y_{\mathrm{min}}$ and $y=y_{\mathrm{max}}$. Another
possibility is the modification of equation \eref{eq:sde-y-2} to get rapidly
decreasing steady-state PDF when the stochastic variable $y$ acquires values
outside of the interval $[y_{\mathrm{min}},y_{\mathrm{max}}]$. For example, the
steady-state PDF 
\begin{equation}
P_{0}(x,y)\sim p(x)y^{-\lambda}\exp\left\{ -\left(\frac{y_{\mathrm{min}}}{y}\right)^{m}
-\left(\frac{y}{y_{\mathrm{max}}}\right)^{m}\right\} 
\end{equation}
with $m>0$ has a power-law dependence on $y$ when $y_{\mathrm{min}}\ll y\ll
y_{\mathrm{max}}$ and exponential cut-offs when $y$ is outside of the interval
$[y_{\mathrm{min}},y_{\mathrm{max}}]$. This exponentially restricted
steady-state PDF is a result of the SDE
\begin{equation}
\rmd y_{t}=\sigma^{2}\left(\eta+1-\frac{\lambda}{2}+\frac{m}{2}
\left(\frac{y_{\mathrm{min}}^{m}}{y_{t}^{m}}
-\frac{y_{t}^{m}}{y_{\mathrm{max}}^{m}}\right)\right)y_{t}^{2\eta+1}\rmd t
+\sigma y_{t}^{\eta+1}\rmd W_{t}^{\prime}
\label{eq:restricted-1}
\end{equation}
obtained from equation \eref{eq:sde-y-2} by introducing additional terms
in the drift.

\subsection{Limiting frequencies}

The restriction of the diffusion of the stochastic variable $y$ to the interval
$y_{\mathrm{min}}\ll y\ll y_{\mathrm{max}}$ makes the scaling
\eref{eq:transition-scaling} only approximate. As a result, the power-law part
of the PSD is limited to a finite range of frequencies $f_{\mathrm{min}}\ll f\ll
f_{\mathrm{max}}$. Let us estimate the limiting frequencies $f_{\mathrm{min}}$
and $f_{\mathrm{max}}$. The limiting values $y=y_{\mathrm{min}}$ and
$y=y_{\mathrm{max}}$ should also participate in the scaling and
equation \eref{eq:transition-scaling} for the transition probability corresponding
to SDEs \eref{eq:sde-x-2} and \eref{eq:sde-y-2} becomes
\begin{equation}
aP(x',ay,t|x,ay,0;ay_{\mathrm{min}},ay_{\mathrm{max}})=P(x',y',a^{\mu}t|,x,y,0;y_{\mathrm{min}},
y_{\mathrm{max}})\,.\label{eq:transition-scaling-bound}
\end{equation}
Here $y_{\mathrm{min}}$, $y_{\mathrm{max}}$ enter as parameters of the
transition probability. Similarly, the steady-state PDF
$P_{0}(x,y;y_{\mathrm{min}},y_{\mathrm{max}})$ has the scaling property
\begin{equation}
aP_{0}(x,ay;ay_{\mathrm{min}},ay_{\mathrm{max}})=P_{0}(x,y;y_{\mathrm{min}},y_{\mathrm{max}})\,.
\label{eq:steady-scaling-bound}
\end{equation}
Inserting equations \eref{eq:transition-scaling-bound} and \eref{eq:steady-scaling-bound}
into \eref{eq:autocorr} we get
\begin{equation}
C(t,ay_{\mathrm{min}},ay_{\mathrm{max}})=C(a^{\mu}t,y_{\mathrm{min}},y_{\mathrm{max}})\,.
\end{equation}
From this scaling of the autocorrelation function it follows that time $t$ should
enter only in combinations with the limiting values $y_{\mathrm{min}}t^{1/\mu}$
and $y_{\mathrm{max}}t^{1/\mu}$. We can expect that the influence of the
limiting values can be neglected and the scaling \eref{eq:transition-scaling}
be approximately valid when $y_{\mathrm{min}}t^{1/\mu}\ll1$ and
$y_{\mathrm{max}}t^{1/\mu}\gg1$. In other words, we expect that the scaling
\eref{eq:transition-scaling} holds when time $t$ is in the interval
$\sigma^{-2}y_{\mathrm{max}}^{-\mu}\ll t\ll\sigma^{-2}y_{\mathrm{min}}^{-\mu}$
when $\mu>0$ and in the interval $\sigma^{-2}y_{\mathrm{min}}^{-\mu}\ll
t\ll\sigma^{-2}y_{\mathrm{max}}^{-\mu}$ when $\mu<0$. Using
equation \eref{eq:wiener-khintchine} the frequency range where the PSD has
$1/f^{\beta}$ behavior can be estimated as
\begin{eqnarray}
\sigma^{2}y_{\mathrm{min}}^{\mu} \ll 2\pi f\ll\sigma^{2}y_{\mathrm{max}}^{\mu}\,,\qquad\mu>0
\label{eq:freq-range}\\
\sigma^{2}y_{\mathrm{max}}^{\mu} \ll 2\pi f\ll\sigma^{2}y_{\mathrm{min}}^{\mu}\,,\qquad\mu<0
\end{eqnarray}
We see that the width of the frequency range where the PSD has $1/f^{\beta}$
behavior grows with increase of the ratio $y_{\mathrm{max}}/y_{\mathrm{min}}$.
For $\mu=0$ (which corresponds to $\eta=0$) the width of the frequency region
\eref{eq:freq-range} is zero and we do not have $1/f^{\beta}$ power spectral
density.

\section{Numerical approach}

\label{sec:numer} Since analytical solution of stochastic differential equations
can be obtained only in particular cases, there is a need of numerical solution.
Using Euler-Maruyama method with small time step $\Delta t$ for numerical
solution of SDEs \eref{eq:sde-x-2} and \eref{eq:sde-y-2}, we get the discretized
equations 
\begin{eqnarray}
x_{k+1} = x_{k}+a(x_{k})y_{k}^{2\eta}\Delta t
+b(x_{k})y_{k}^{\eta}\sqrt{\Delta t}\varepsilon_{k}\,,\\
y_{k+1} = y_{k}+\sigma^{2}\left(\eta+1-\frac{\lambda}{2}\right)y_{k}^{2\eta+1}\Delta t
+\sigma y_{k}^{\eta+1}\sqrt{\Delta t}\xi_{k}\,.
\end{eqnarray}
Here $\varepsilon_{k}$ and $\xi_{k}$ are independent random variables with the 
standard normal distribution. However, for numerical solution of nonlinear
equations the solution schemes involving a fixed time step $\Delta t$ can be
inefficient. For example, in equations \eref{eq:sde-x-2} and \eref{eq:sde-y-2} with
$\eta>0$, large values of stochastic variable $y$ lead to large coefficients and
thus require a very small time step. The numerical solution scheme can by
improved by using a variable time step that becomes small only when $y$ becomes
large. Such method of solution of a single nonlinear SDE has been proposed in
\cite{Kaulakys2004,Ruseckas2015}. The variable time step is equivalent to the
introduction of the internal time $\tau$ that is different from the real,
physical, time $t$ \cite{Ruseckas2015}.

In order to make the solution more efficient we introduce an internal,
operational, time $\tau$ by the equation
\begin{equation}
\rmd\tau_{t}=y_{t}^{2\eta}\rmd t\,.\label{eq:internal-time}
\end{equation}
We assume that the zero of the internal time $\tau$ coincides with the zero of
the physical time $t$, thus the initial condition for the internal time is
$\tau_{t=0}=0$. Since $y_{t}>0$, from equation \eref{eq:internal-time} it follows
that $\tau_{t}$ is a strictly increasing function of time $t$. Let us obtain the
SDEs for the stochastic variables $x$ and $y$ in the internal time $\tau$. To do
this we proceed similarly as in \cite{Ruseckas2015} and consider the joint
PDF $P_{x,y,\tau}(x,y,\tau;t)$ of the stochastic variables $x$, $y$ and $\tau$.
The PDF $P(x,y;t)$ can be calculated using the equation
\begin{equation}
P_{x,y}(x,y,t)=\int P_{x,\tau}(x,y,\tau;t)\,\rmd\tau\,.
\end{equation}
Equations \eref{eq:sde-x-2}, \eref{eq:sde-y-2}, and \eref{eq:internal-time}
lead to the Fokker-Planck equation for the PDF $P_{x,y,\tau}(x,y,\tau;t)$
\begin{eqnarray}
\fl\frac{\partial}{\partial t}P_{x,y,\tau} = -y^{2\eta}\frac{\partial}{\partial x}a(x)
P_{x,y,\tau}-\sigma^{2}\left(\eta+1-\frac{\lambda}{2}\right)
\frac{\partial}{\partial y}y^{2\eta+1}P_{x,y,\tau}
-y^{2\eta}\frac{\partial}{\partial\tau}P_{x,y,\tau}\nonumber \\
+\frac{1}{2}y^{2\eta}\frac{\partial^{2}}{\partial x^{2}}b(x)^{2}P_{x,y,\tau}
+\frac{1}{2}\sigma^{2}\frac{\partial^{2}}{\partial y^{2}}y^{2\eta+2}P_{x,y,\tau}\,.
\label{eq:FP-x-y-tau}
\end{eqnarray}
Since the zero of the internal time $\tau$ coincides with the zero
of the physical time $t$, the initial condition for equation \eref{eq:FP-x-y-tau}
is $P_{x,y,\tau}(x,y,\tau;0)=P(x,y;0)\delta(\tau)$. Matching of the zeros
of $\tau$ and $t$ leads also to the boundary condition $P_{x,y,\tau}(x,y,0;t)=0$
for $t>0$, because $\tau$ and t are strictly increasing.

Instead of $x$, $y$ and $\tau$ we can consider $x$, $y$ and $t$
as stochastic variables. The physical time $t$ is related to the
operational time $\tau$ via equation \eref{eq:internal-time}, therefore,
the joint PDF $P_{x,y,t}(x,y,t;\tau)$ of the stochastic variables
$x$, $y$ and $t$ is related to the PDF $P_{x,y,\tau}(x,y,\tau;t)$
according to the equation
\begin{equation}
P_{x,y,t}(x,y,t;\tau)=y^{2\eta}P_{x,y,\tau}(x,y,\tau;t)\,.\label{eq:p-x-y-t}
\end{equation}
Another way to get this relation is to notice that the third term
on the right hand side of equation \eref{eq:FP-x-y-tau} contains the
derivative $\frac{\partial}{\partial\tau}$ and thus should be equal
to $-\frac{\partial}{\partial\tau}P_{x,y,t}$. Inserting \eref{eq:p-x-y-t}
into equation \eref{eq:FP-x-y-tau} we get
\begin{eqnarray}
\fl\frac{\partial}{\partial\tau}P_{x,y,t} = -\frac{\partial}{\partial x}a(x)P_{x,y,t}
-\sigma^{2}\left(\eta+1-\frac{\lambda}{2}\right)\frac{\partial}{\partial y}yP_{x,y,t}
-\frac{\partial}{\partial t}\frac{1}{y^{2\eta}}P_{x,y,t}\nonumber\\
+\frac{1}{2}\frac{\partial^{2}}{\partial x^{2}}b(x)^{2}P_{x,y,t}
+\frac{1}{2}\sigma^{2}\frac{\partial^{2}}{\partial y^{2}}y^{2}P_{x,y,t}\,.
\label{eq:FP-x-y-t}
\end{eqnarray}
The initial condition for equation \eref{eq:FP-x-y-t} is $P_{x,y,t}(x,t;0)=P(x,y;0)\delta(t)$.
In addition, there is a boundary condition $P_{x,y,t}(x,y,0;\tau)=0$
for $\tau>0$. The Fokker-Planck equation \eref{eq:FP-x-y-t} can
be obtained from the coupled SDEs
\begin{eqnarray}
\rmd x_{\tau} = a(x_{\tau})\rmd\tau+b(x_{\tau})\rmd W_{\tau}\,,\label{eq:sde-x-inter}\\
\rmd y_{\tau} = \sigma^{2}\left(\eta+1-\frac{\lambda}{2}\right)y_{\tau}\rmd\tau
+\sigma y_{\tau}\rmd W_{\tau}^{\prime}\,,\label{eq:sde-y-inter}\\
\rmd t_{\tau} = \frac{1}{y_{\tau}^{2\eta}}\rmd\tau\,.
\end{eqnarray}
Discretizing the internal time $\tau$ with the step $\Delta\tau$
and using the Euler-Maruyama approximation for SDEs \eref{eq:sde-x-inter} 
and \eref{eq:sde-y-inter}, we get
\begin{eqnarray}
x_{k+1} = x_{k}+a(x_{k})\Delta\tau+b(x_{k})\sqrt{\Delta\tau}\varepsilon_{k}\,,
\label{eq:discr-x}\\
y_{k+1} = y_{k}+\sigma^{2}\left(\eta+1-\frac{\lambda}{2}\right)y_{k}\Delta\tau
+\sigma y_{k}\sqrt{\Delta\tau}\xi_{k}\,,\label{eq:discr-y}\\
t_{k+1} = t_{k}+\frac{\Delta\tau}{y_{k}^{2\eta}}\,.\label{eq:discr-t}
\end{eqnarray}
Equations \eref{eq:discr-x}--\eref{eq:discr-t} provide the numerical method for
solving coupled SDEs~\eref{eq:sde-x-2} and \eref{eq:sde-y-2}. One can interpret
equations \eref{eq:discr-x}--\eref{eq:discr-t} as an Euler-Maruyama scheme with a
variable time step $\Delta t_{k}=\Delta\tau/y_{k}^{2\eta}$ that adapts to the
coefficients in the SDEs. As a consequence of the introduction of the internal
time the increments of the real, physical, time $t$ become random. To get the
discretization of time with fixed steps the signal generated in such a way
should be interpolated.

\begin{figure}
\includegraphics[width=0.45\textwidth]{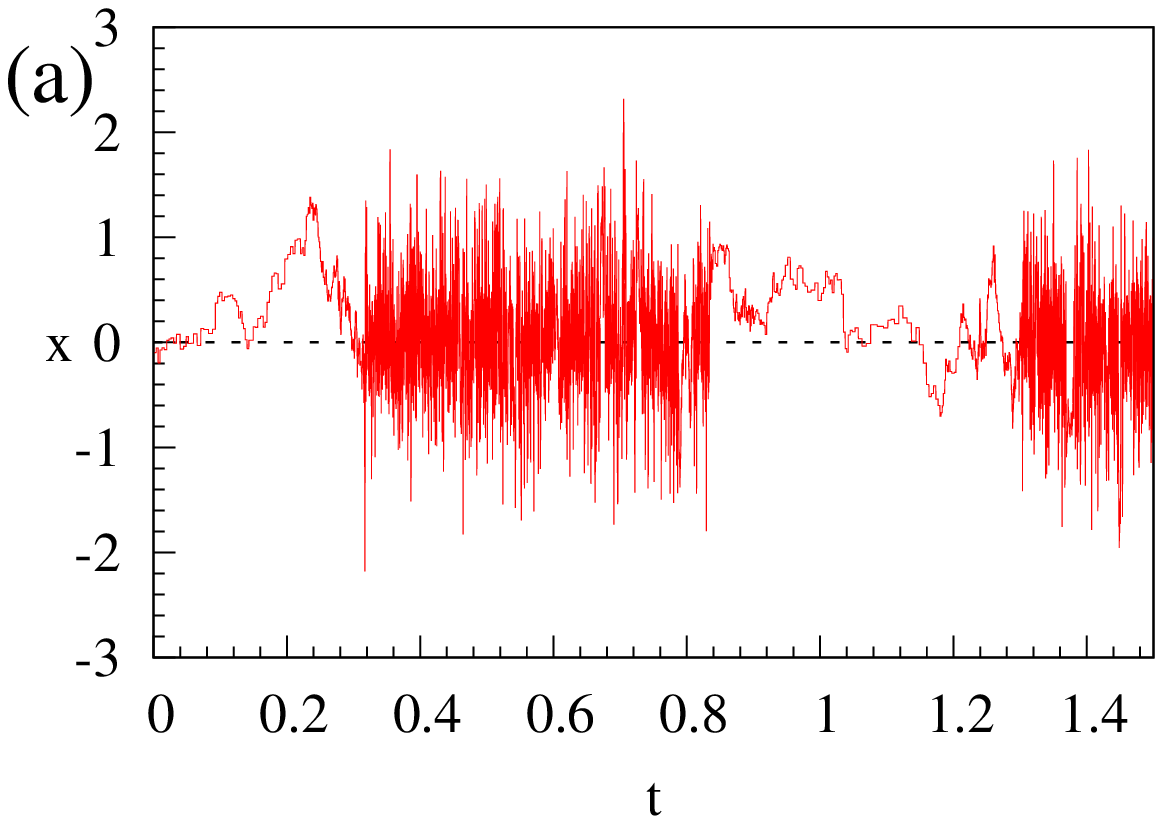}\includegraphics[width=0.45\textwidth]{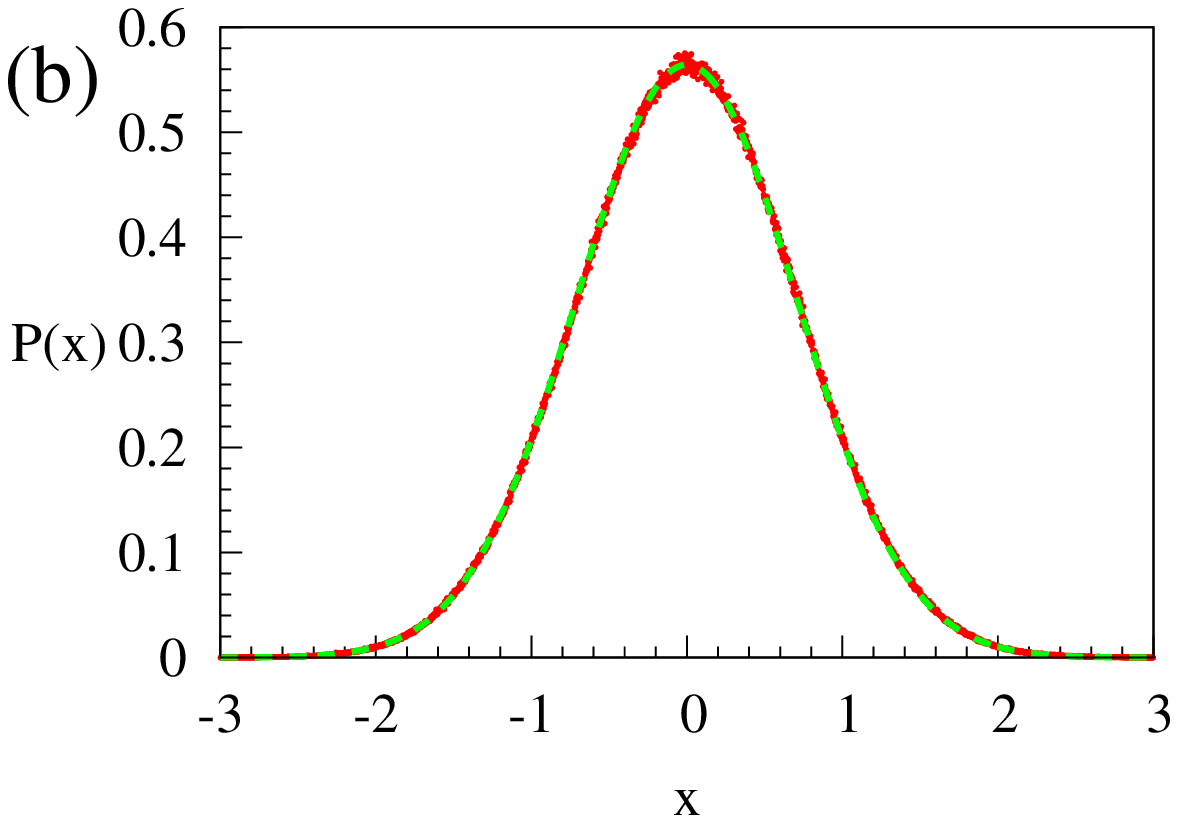}\\
\includegraphics[width=0.45\textwidth]{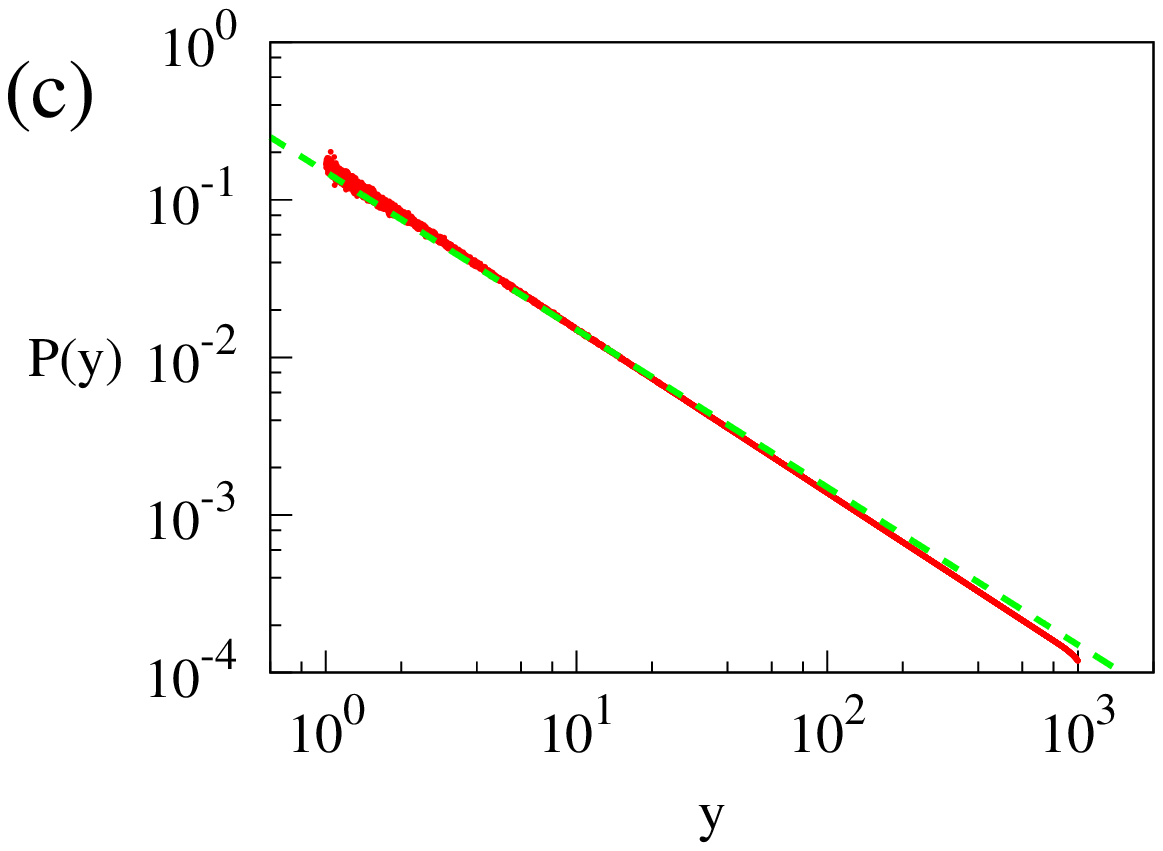}\includegraphics[width=0.45\textwidth]{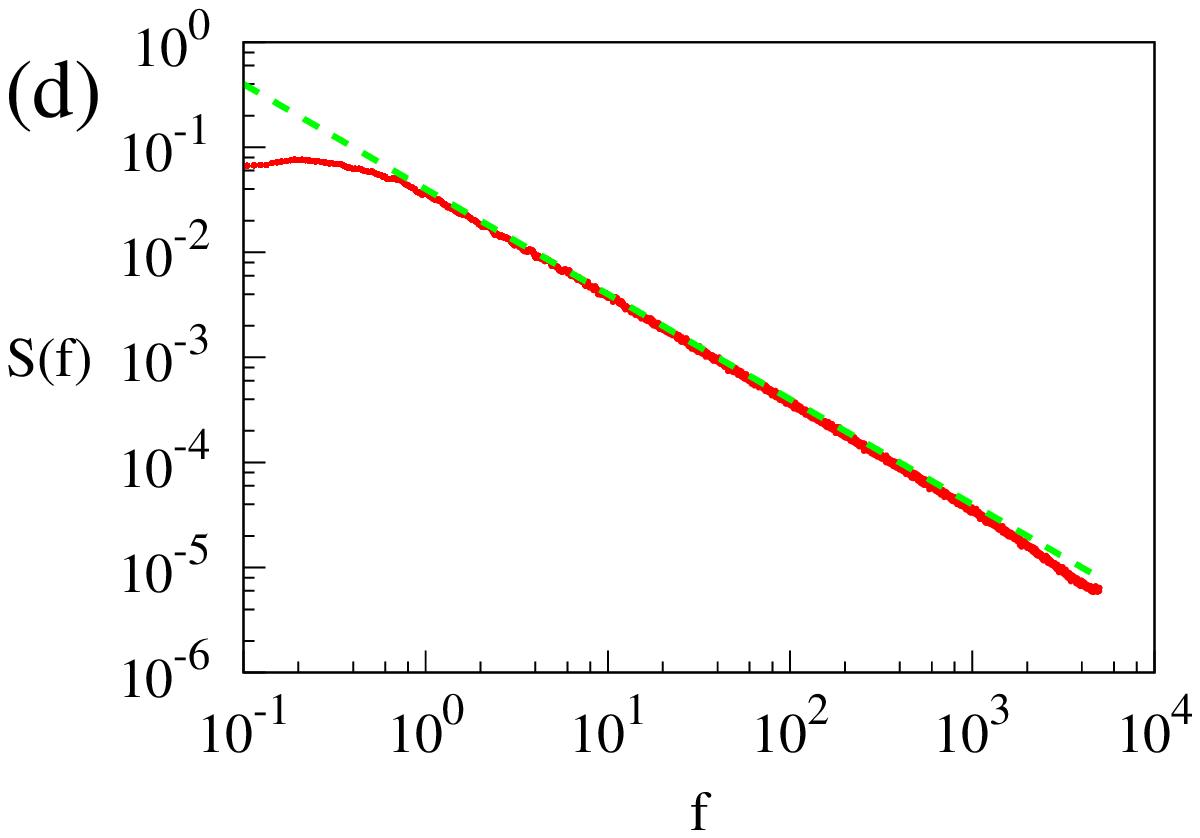}
\caption{(a) Typical signal $x$ generated by equations \eref{eq:sde-numer-x} and
\eref{eq:sde-numer-y}. Reflective boundaries at $y_{\mathrm{min}}$ and
$y_{\mathrm{max}}$ have been used for equation \eref{eq:sde-numer-y}. (b) The PDF of
the signal intensity. The dashed (green) line shows the Gaussian curve. (c) The
PDF of the stochastic variable $y$. The dashed (green) line shows the power-law
with the exponent $-1$. (d) The PSD of the signal $x$. The dashed (green) line
shows the slope $f^{-1}$. Used parameters are $\eta=1$, $\lambda=1$,
$y_{\mathrm{min}}=1$, $y_{\mathrm{max}}=1000$, $\gamma=1$ and $\sigma=1$.}
\label{fig:numer}
\end{figure}

As an example, let us solve the equations
\begin{eqnarray}
\rmd x_{t} = -\gamma y_{t}^{2\eta}x_{t}\rmd t+y_{t}^{\eta}\rmd W_{t}\,,
\label{eq:sde-numer-x}\\
\rmd y_{t} = \sigma^{2}\left(\eta+1-\frac{\lambda}{2}\right)y_{t}^{2\eta+1}\rmd t
+\sigma y_{t}^{\eta+1}\rmd W_{t}^{\prime}\,.
\label{eq:sde-numer-y}
\end{eqnarray}
For the stochastic variable $y$ we assume reflective boundaries at
$y=y_{\mathrm{min}}$ and $y=y_{\mathrm{max}}$. In this case the
coefficients $a(x)$ and $b(x)$ in equation \eref{eq:sde-x-2} are $a(x)=-\gamma x$
and $b(x)=1$, leading to the Gaussian steady-state PDF of $x$, 
\begin{equation}
p(x)=\sqrt{\frac{\gamma}{\pi}}\rme^{-\gamma x^{2}}\,.
\end{equation}
The quantity $y^{2\eta}$ in equation \eref{eq:sde-numer-x} represents a fluctuating
relaxation rate.

Comparison of the numerically obtained steady state PDF and the PSD with
analytical expressions for the system of SDEs \eref{eq:sde-numer-x} and
\eref{eq:sde-numer-y} with $\eta=1$ and $\lambda=1$ is presented in
\fref{fig:numer}. Typical signal $x_{t}$ generated by equations
\eref{eq:sde-numer-x} and \eref{eq:sde-numer-y} is shown in \fref{fig:numer}(a).
As one can see, the signal exhibits a structure consisting of the periods of
slow and fast fluctuations. The fast fluctuations correspond to the peaks or
bursts of the stochastic variable $y$. Note, that due to large difference
between slowest and fastest fluctuation rates the signal in the periods of fast
fluctuations in \fref{fig:numer}(a) visually resembles white noise. However, the
actual signal changes according to SDE \eref{eq:sde-numer-x}, the periods of
fast fluctuations are similar to the periods of slow fluctuations compressed in
time. Analysis of nonlinear SDEs similar to \eref{eq:sde-numer-y}, performed in
\cite{Kaulakys2009}, reveals that the sizes of the bursts are approximately
proportional to the squared durations of the bursts. The distributions of burst
and inter-burst durations have power-law parts, with the numerically estimated
power-law exponent of the PDF of the inter-burst durations approximately equal
to $-3/2$. Intermittent behavior, similar to the behavior shown in
\fref{fig:numer}(a), can be connected with $1/f$ noise. For example, it is known
that intermittent behavior in iterative maps at the edge of chaos can lead to
$1/f$ noise \cite{Schuster1988}.
In figures \ref{fig:numer}(b) and \ref{fig:numer}(c) we see a good agreement of
the numerically calculated steady-state PDFs of the stochastic variables $x$ and
$y$ with the analytical expressions. The PSD of the signal $x_{t}$ is shown in
\fref{fig:numer}(d). Numerical solution of the equations confirms the presence
of the frequency region for which the power spectral density has $1/f^{\beta}$
dependence with $\beta=1$.

\section{More general form of equations} 

\label{sec:general}Coupled nonlinear SDEs~\eref{eq:sde-x-2} and
\eref{eq:sde-y-2} exhibit the separation between the magnitude of the
fluctuations of the signal $x_{t}$ and the rate of fluctuations. The
steady-state PDF of the signal is determined only by the coefficients $a(x)$ and
$b(x)$ in equation \eref{eq:sde-x-2}, whereas equation \eref{eq:sde-y-2} describes the
fluctuating rate that does not depend on the signal. However,
equations \eref{eq:sde-x-2} and \eref{eq:sde-y-2} are not the most general form of
coupled SDEs that are allowed by scaling properties required to get
$1/f^{\beta}$ spectrum. For completeness, in this section we will consider a more
general form of equations.

In general, scaling of time $t$ in the transition probability can lead to
scaling of both $x$ and $y$, therefore instead of
equation \eref{eq:transition-scaling} in this section we will consider a more
general scaling property of the transition probability, 
\begin{equation}
a^{\rho+1}P(a^{\rho}x',ay,t|a^{\rho}x,ay,0)=P(x',y',a^{\mu}t|,x,y,0)\,.
\label{eq:transition-scaling-2}
\end{equation}
We also assume scaling property of the steady-state PDF similar to
the scaling property \eref{eq:transition-scaling-2} of the transition
probability 
\begin{equation}
P_{0}(a^{\rho}x,ay)\sim a^{-\lambda}P_{0}(x,y)\,.\label{eq:steady-scaling-2}
\end{equation}
Here $\mu$, $\rho$ and $\lambda$ are the scaling exponents. From
equation \eref{eq:steady-scaling-2} it follows that the steady-state PDF should have
the form
\begin{equation}
P_{0}(x,y)=p(xy^{-\rho})y^{-\lambda}\,,\label{eq:steady-form}
\end{equation}
where $p(\cdot)$ is an arbitrary function. Using equations \eref{eq:autocorr},
\eref{eq:transition-scaling-2}, and \eref{eq:steady-scaling-2} and performing
a change of variables we obtain
\begin{eqnarray}
\fl C(at) & = \int \rmd x\rmd y\int \rmd x'\rmd y'xx'P_{0}(x,y)P(x',y',at|x,y,0)\\
\fl & \sim \int \rmd x\rmd y\int \rmd x'\rmd y'xx'a^{\frac{\lambda}{\mu}}
P_{0}(a^{\frac{\rho}{\mu}}x,a^{\frac{1}{\mu}}y)a^{\frac{\rho+1}{\mu}}
P(a^{\frac{\rho}{\mu}}x',a^{\frac{1}{\mu}}y,t|a^{\frac{\rho}{\mu}}x,a^{\frac{1}{\mu}}y,0)\\
\fl & \sim a^{\frac{\lambda-1-3\rho}{\mu}}\int \rmd u\rmd v\int \rmd u'\rmd v'uu'P_{0}(u,v)P(u',v,t|u,v,0)\,.
\end{eqnarray}
Therefore, the autocorrelation function has the scaling property
\eref{eq:autocorr-scaling} required to get $1/f^{\beta}$ PSD, with the exponent
$\beta$ given by equation
\begin{equation}
\beta=1+\frac{\lambda-1-3\rho}{\mu}\,.\label{eq:beta-3}
\end{equation}
In this case we obtain pure $1/f$ noise when $\lambda=1+3\rho$.

To get the scaling property \eref{eq:transition-scaling-2} of the transition
probability, we will consider the following coupled It\^o SDEs 
\begin{eqnarray}
\rmd x_{t} = a(x_{t}y_{t}^{-\rho})y_{t}^{2\eta+\rho}\rmd t
+b(x_{t}y_{t}^{-\rho})y_{t}^{\eta+\rho}\rmd W_{t}\,,\label{eq:sde-x-3}\\
\rmd y_{t} = f(x_{t}y_{t}^{-\rho})y_{t}^{2\eta+1}\rmd t
+g(x_{t}y_{t}^{-\rho})y_{t}^{\eta+1}\rmd W_{t}^{\prime}\,.
\label{eq:sde-y-3}
\end{eqnarray}
Here $W_{t}$ and $W_{t}^{\prime}$ are standard Wiener processes. Note, that
equations \eref{eq:sde-x-3} and \eref{eq:sde-y-3} do not have the most general form
compatible with the scaling property \eref{eq:steady-scaling-2}, because in
general both noises $W_{t}$ and $W_{t}^{\prime}$ can affect both stochastic
variables $x$ and $y$. However, for simplicity we will not consider the most
general case. One can see that equations \eref{eq:sde-x-3} and \eref{eq:sde-y-3}
indeed lead to the scaling of transition probability
\eref{eq:transition-scaling-2}. Changing the variables $x$ and $y$ in
equations \eref{eq:sde-x-3} and \eref{eq:sde-y-3} to the scaled variables
$x_{\mathrm{s}}=a^{\rho}x$ and $y_{\mathrm{s}}=ay$ or introducing the scaled
time $t_{\mathrm{s}}=a^{2\eta t}$ and taking into account the property of the
Wiener process $\rmd W_{t_{\mathrm{s}}}\stackrel{d}{=}a^{\eta}\rmd W_{t}$, we get the
same resulting equations. Therefore, the change of the time scale is equivalent
to the corresponding change of scale of the variables $x$ and $y$, according to
equation \eref{eq:transition-scaling-2} with the scaling exponent $\mu=2\eta$. 

The connection between the coefficients $f(\cdot)$ and $g(\cdot)$
we will determine by requiring the steady-state PDF of the form \eref{eq:steady-form}.
The Fokker-Planck equation corresponding to the SDEs \eref{eq:sde-x-3} and 
\eref{eq:sde-y-3} is
\begin{eqnarray}
\fl\frac{\partial}{\partial t}P = -y^{2\eta+\rho}\frac{\partial}{\partial x}
a(xy^{-\rho})P -\frac{\partial}{\partial y}f(xy^{-\rho})y^{2\eta+1}P\nonumber\\
+\frac{1}{2}y^{2\eta+2\rho}\frac{\partial^{2}}{\partial x^{2}}b^{2}(xy^{-\rho})P
+\frac{1}{2}\frac{\partial^{2}}{\partial y^{2}}g^{2}(xy^{-\rho})y^{2\eta+2}P\,,
\end{eqnarray}
therefore, the steady-state PDF $P_{0}(x,y)$ is the solution of the
equation
\begin{eqnarray}
\fl -y^{2\eta+\rho}\frac{\partial}{\partial x}a(xy^{-\rho})P
-\frac{\partial}{\partial y}f(xy^{-\rho})y^{2\eta+1}P
+\frac{1}{2}y^{2\eta+2\rho}\frac{\partial^{2}}{\partial x^{2}}b^{2}(xy^{-\rho})P\nonumber\\
+\frac{1}{2}\frac{\partial^{2}}{\partial y^{2}}g^{2}(xy^{-\rho})y^{2\eta+2}P=0\,.
\label{eq:stationary-FP-2}
\end{eqnarray}
\Eref{eq:stationary-FP-2} can be written in terms of the components
of the probability current
\begin{eqnarray}
J_{x}(x,y) = y^{2\eta+\rho}a(xy^{-\rho})P_{0}-\frac{1}{2}y^{2\eta+2\rho}
\frac{\partial}{\partial x}b^{2}(xy^{-\rho})P_{0}\,,\label{eq:sx-2}\\
J_{y}(x,y) = f(xy^{-\rho})y^{2\eta+1}P_{0}
-\frac{1}{2}\frac{\partial}{\partial y}g^{2}(xy^{-\rho})y^{2\eta+2}P_{0}\,.
\label{eq:sy-2}
\end{eqnarray}
Inserting steady-state PDF \eref{eq:steady-form} into equations \eref{eq:sx-2} and
\eref{eq:sy-2} we get
\begin{eqnarray}
\fl J_{x}(x,y) = y^{2\eta+\rho-\lambda}\left[a(xy^{-\rho})p(xy^{-\rho})
-\frac{1}{2}y^{\rho}\frac{\partial}{\partial x}b^{2}(xy^{-\rho})p(xy^{-\rho})\right]\,,
\label{eq:sx-3}\\
\fl J_{y}(x,y) = y^{2\eta+1-\lambda}g^{2}(xy^{-\rho})p(xy^{-\rho})\nonumber\\
\times\left[\frac{f(xy^{-\rho})}{g^{2}(xy^{-\rho})}-\eta-1+\frac{\lambda}{2}
+\rho xy^{-\rho}\left(\frac{g^{\prime}(xy^{-\rho})}{g(xy^{-\rho})}
+\frac{1}{2}\frac{p^{\prime}(xy^{-\rho})}{p(xy^{-\rho})}\right)\right]\,.
\label{eq:sy-3}
\end{eqnarray}
Assuming that the $x$-component of the probability current $J_{x}$
should vanish at the boundaries that are not parallel to $x$ axis,
we get that the expression in the square brackets in equation \eref{eq:sx-3}
should be zero for different values of $y$. Therefore, the function
$p(\cdot)$ should be a solution of the differential equation
\begin{equation}
a(z)p(z)-\frac{1}{2}\frac{\rmd}{\rmd z}b^{2}(z)p(z)=0\,.\label{eq:p-z}
\end{equation}
This equation means that the function $p(\cdot)$ in equation \eref{eq:steady-form}
is determined only by the coefficients of equation \eref{eq:sde-x-3}.
Similarly, assuming that the $y$-component of the probability current
$J_{y}$ should vanish at the boundaries that are not parallel to
$y$ axis we get that the expression in the square brackets in equation \eref{eq:sy-3}
should be zero for different values of $y$. Therefore, the coefficient
$f(\cdot)$ is related to the coefficients $a(\cdot)$, $b(\cdot)$
and $g(\cdot)$ via the equation
\begin{equation}
f(z)=\left[\eta+1-\frac{\lambda}{2}-\rho z\left(\frac{g^{\prime}(z)}{g(z)}
+\frac{1}{2}\frac{p^{\prime}(z)}{p(z)}\right)\right]g^{2}(z)\,.
\label{eq:f-g}
\end{equation}

Let us consider some particular choices of the coefficients $f(\cdot)$
and $g(\cdot)$ in equation \eref{eq:sde-y-3}. According to equation \eref{eq:f-g},
constant coefficient $g(z)=\sigma=\mathrm{const}$ leads to
\begin{equation}
f(z)=\sigma^{2}\left[\eta+1-\frac{\lambda}{2}-\rho z\left(\frac{a(z)}{b^{2}(z)}
-\frac{b^{\prime}(z)}{b(z)}\right)\right]\,.
\label{eq:f-const-g}
\end{equation}
Here we used equation \eref{eq:p-z} for the function $p(z)$. When
\begin{equation}
\frac{g^{\prime}(z)}{g(z)}+\frac{1}{2}\frac{p^{\prime}(z)}{p(z)}=0\,,
\label{eq:g-1}
\end{equation}
from equation \eref{eq:f-g} it follows that the stochastic variable $x$ enters into
the coefficients $f(\cdot)$ and $g(\cdot)$ only as an argument of the function
$p(\cdot)$. The solution of equation \eref{eq:g-1} is $g(z)=\sigma p(z)^{-1/2}$.
Consequently, $f(z)=\sigma^{2}(\eta+1-\lambda/2)p(z)^{-1}$ and
equations \eref{eq:sx-3} and \eref{eq:sy-3} take the form
\begin{eqnarray}
\rmd x_{t} = a(x_{t}y_{t}^{-\rho})y_{t}^{2\eta+\rho}\rmd t
+b(x_{t}y_{t}^{-\rho})y_{t}^{\eta+\rho}\rmd W_{t}\,,\\
\rmd y_{t} = \sigma^{2}\left(\eta+1-\frac{\lambda}{2}\right)
\frac{y_{t}^{2\eta+1}}{p(x_{t}y_{t}^{-\rho})}\rmd t
+\frac{\sigma y_{t}^{\eta+1}}{\sqrt{p(x_{t}y_{t}^{-\rho})}}\rmd W_{t}^{\prime}\,.
\end{eqnarray}

As an example, let us take the SDE \eref{eq:sde-x-3} describing
the fluctuations of the signal $x$, 
\begin{equation}
\rmd x_{t}=-y_{t}x_{t}\rmd t+y_{t}^{\nu}\rmd W_{t}\,.\label{eq:example-2-x}
\end{equation}
The stochastic variable $y$ in equation \eref{eq:example-2-x} represents a
fluctuating relaxation rate. The value of $\nu=\frac{1}{2}$ corresponds to the
fluctuation-dissipation theorem. However, there are some cases where the
fluctuation-dissipation theorem cannot be applied and other values of $\nu$ are
possible. The violation of the fluctuation-dissipation theorem has been found in
the finite dimensional spin glasses \cite{Marinari1998} and in the systems out
of equilibrium \cite{Lobaskin2006}. The theoretical study of motion of colloidal
particles being confined in a harmonic well and dragged by a shear flow also
shows violation of the fluctuation-dissipation theorem \cite{Mauri2006}.
Comparing equation \eref{eq:example-2-x} with equation \eref{eq:sde-x-3} we have
$a(z)=-z$, $b(z)$, $\eta=\frac{1}{2}$, $\rho=\nu-\frac{1}{2}$. Using
equations \eref{eq:sde-y-3} and \eref{eq:f-const-g} we obtain the second equation
\begin{equation}
\rmd y_{t}=\sigma^{2}\left(\frac{3}{2}-\frac{\lambda}{2}
+\left(\nu-\frac{1}{2}\right)y_{t}^{1-2\nu}x_{t}^{2}\right)y_{t}^{2}\rmd t
+\sigma y_{t}^{\frac{3}{2}}\rmd W_{t}^{\prime}\,.
\label{eq:example-2-y}
\end{equation}
According to \eref{eq:beta-3}, equations \eref{eq:example-2-x} and 
\eref{eq:example-2-y} generate the signal $x_{t}$ with power-law
behavior $1/f^{\beta}$ of the PSD in a wide range of frequencies,
with the exponent $\beta=\lambda+3\left(\frac{1}{2}-\nu\right)$.

As an another example let us consider the SDE \eref{eq:sde-x-3} with the coefficients
$a(z)=0$ and $b(z)=\mathrm{const}$:
\begin{equation}
\rmd x_{t}=b y_{t}^{\eta+\rho}\rmd W_{t}\,.\label{eq:example-3-x}
\end{equation}
To get stationary solution of the corresponding Fokker-Planck equation, equation
\eref{eq:example-3-x} should be taken together with boundaries limiting the
region of diffusion of stochastic variable $x$. For such coefficients $a(z)$ and
$b(z)$ the solution of equation \eref{eq:p-z} is $p(z)=\mathrm{const}$.
Equations \eref{eq:sde-y-3} and \eref{eq:f-const-g} yield the second SDE
\begin{equation}
\rmd y_{t}=\sigma^{2}\left(\eta +1-\frac{\lambda}{2}\right)y_{t}^{2\eta+1}\rmd t
+\sigma y_{t}^{\eta+1}\rmd W_{t}^{\prime}\,.
\label{eq:example-3-y}
\end{equation}
We see that in this case the second equation does not depend on $x$.

\section{Discussion and conclusions}

\label{sec:concl}
Coupled Langevin equations have been used to describe many physical phenomena.
For example, hot-carrier transport in semiconductors has been modeled by
linearly coupled Langevin equations \cite{Kuhn1992}; nonlinear coupled Langevin
equations have been used to study pressure time series \cite{Lind2005}. One
nonlinear SDE with fluctuating parameter can be interpreted as a pair of coupled
SDEs \cite{Jizba2008}. Equations with time varying parameter being a Gaussian
colored noise (Ornstein-Uhlenbeck process) have been used to model wind farm
power production output dependence on wind velocity \cite{Milan2014} and
atmospheric turbulence in radio signal detection \cite{Kloeden1992}. In this
paper we study nonlinear SDEs where the fluctuating parameter enters both
diffusion and drift coefficients as a power-law function.

Coupled SDEs are also used in finance and econophysics for stochastic volatility
models \cite{Slanina2014}, some of those models correspond to equations
presented in \sref{sec:general}. For example, SDE \eref{eq:example-3-x} and
SDE \eref{eq:example-3-y} with an additional drift term causing exponential
restriction of the steady-state PDF, when the parameters $\eta$ and $\rho$ take
values $\eta = -\frac{1}{2}$, $\rho = 1$ have the form of the Heston model
\cite{Heston1993}
\begin{eqnarray}
\rmd x_{t} = \sqrt{y_{t}}\rmd W_{t}\,,\label{eq:heston-x}\\
\rmd y_{t} = \frac{1}{2}\sigma^{2}\left(1-\lambda -\frac{y_{t}}{y_{\mathrm{max}}}\right)
\rmd t +\sigma \sqrt{y_{t}}\rmd W_{t}^{\prime}\,.
\label{eq:heston-y}
\end{eqnarray}
In this model the stochastic variable $x$ represents the logarithm of the price
and the stochastic variable $y$ is the volatility.

To illustrate the situation that can be described by the proposed SDEs
\eref{eq:sde-x-2} and \eref{eq:sde-y-2}, let us consider the case with
$\eta=-\frac{1}{2}$. Equations \eref{eq:sde-x-2} and \eref{eq:sde-y-2} then
become
\begin{eqnarray}
\rmd x_{t} = a(x_{t})\frac{1}{y_{t}}\rmd t+b(x_{t})\frac{1}{\sqrt{y_{t}}}\rmd W_{t}\,,
\label{eq:example-4-x}\\
\rmd y_{t} = \frac{1}{2}\sigma^{2}(1-\lambda)\rmd t+\sigma\sqrt{y_{t}}\rmd W_{t}^{\prime}\,.
\label{eq:example-4-y}
\end{eqnarray}
The quantity $y^{-1}$ in equation \eref{eq:example-4-x} has the meaning of the rate
of change, whereas $y$ has the meaning of time interval. According to
equation \eref{eq:beta-3}, the PSD of the signal $x_{t}$ has power-law behavior for
a wide range of frequencies with the power-law exponent
\begin{equation}
\beta=2-\lambda\,.
\end{equation}
We get $1/f$ noise when $\lambda=1$. Assuming that the coefficients
$a(x)$ and $b(x)$ are sufficiently small, we can take $\Delta\tau=1$
in the numerical solution scheme \eref{eq:discr-x}--\eref{eq:discr-t},
leading to the discrete equations 
\begin{eqnarray}
x_{k+1} = x_{k}+a(x_{k})+b(x_{k})\varepsilon_{k}\,,\label{eq:discr-x-3}\\
y_{k+1} = y_{k}\left(1+\frac{1}{2}\sigma^{2}(1-\lambda)+\sigma\xi_{k}\right)\,,
\label{eq:discr-y-3}\\
t_{k+1} = t_{k}+y_{k}\,.\label{eq:discr-t-3}
\end{eqnarray}
In particular, when $\lambda=1$ and the signal $x$ has $1/f$ spectrum,
equation \eref{eq:discr-y-3} becomes $y_{k+1}=y_{k}(1+\sigma\xi_{k})$. We can
interpret equations \eref{eq:discr-x-3}--\eref{eq:discr-t-3} as follows:
equations \eref{eq:discr-y-3} and \eref{eq:discr-t-3} describe a process consisting
of discrete events occurring at time moments $t_{k}$. The inter-event duration is
random and equal to the stochastic variable $y_{k}$. This inter-event duration
slowly changes with time in such a way, that the duration of the next time
interval is equal to the duration of the previous interval multiplied by some
random factor close to $1$. The signal $x_{k}$ changes only during the
occurrence of the events at time moments $t_{k}$ and this change is described by
equation \eref{eq:discr-x-3}.

\Eref{eq:example-4-y} results in the steady-state PDF $P_{0}(y_{t})$ of the
stochastic variable $y_{t}$ having a power-law form with the exponent
$-\lambda$. The PDF $P_k(y_{k})$ of a sequence of $y_{k}$ values generated
according to equation \eref{eq:discr-y-3} differs from $P_{0}(y_{t})$. When $y_{k}$
changes slowly with the index $k$, the PDF $P_k(y_{k})$ should satisfy the
equation $P_{0}(y_{k})\approx\frac{y_{k}}{\langle y_{k}\rangle}P_k(y_{k})$,
because going back from discrete equations to the continuous time one should
assume that each value $y_{k}$ last for the duration also equal $y_{k}$.
Consequently, the PDF $P_k(y_{k})$ is also a power-law with the exponent
$-\lambda^{\prime}$, $\lambda^{\prime}=\lambda+1$. Thus, if $\lambda$ is close
to $1$ then $\lambda^{\prime}$ is close to $2$. 

There are many processes in the nature with the power-law inter-event time
distribution. For example, many human-related activities show power-law decaying
inter-event time distribution with exponents usually varying between $1$ and $2$
\cite{Eckmann2004,Oliveira2005,Dezsoe2006,Vazquez2006}. Power-law distribution
of inter-event times has been observed in neuron-firing sequences
\cite{Kemuriyama2010} and in the timings of earthquakes
\cite{Corral2004,Godano2015}. In addition, power-law decaying inter-event time
distribution is often accompanied by the power-law decaying autocorrelation
function \cite{Karsai2012}.

Let us further assume that the events are due to jumps over the potential
barrier of the height $v$. In many physical systems the escape rate
exponentially depends on the barrier height, therefore we take $y=\rme^{v}$.
Changing the variables in equations \eref{eq:example-4-x} and \eref{eq:example-4-y}
we get the SDEs
\begin{eqnarray}
\rmd x_{t} = a(x_{t})\rme^{-v_{t}}\rmd t+b(x_{t})\rme^{-v_{t}/2}\rmd W_{t}\,,\\
\rmd v_{t} = -\frac{1}{2}\sigma^{2}\lambda \rme^{-v_{t}}\rmd t
+\sigma \rme^{-v_{t}/2}\rmd W_{t}^{\prime}\,.
\end{eqnarray}
Similar to equations \eref{eq:discr-x-3}--\eref{eq:discr-t-3}, numerical
solution scheme with the variable time step $\Delta t_{k}=\rme^{v_{k}}$
yields discrete equations
\begin{eqnarray}
x_{k+1} = x_{k}+a(x_{k})+b(x_{k})\varepsilon_{k}\,,\\
v_{k+1} = v_{k}-\frac{1}{2}\sigma^{2}\lambda+\sigma\xi_{k}\,,
\label{eq:discr-v}\\
t_{k+1} = t_{k}+\rme^{v_{k}}\,.
\end{eqnarray}
From equation \eref{eq:discr-v} we see that the potential $v$ performs a simple
random walk with a constant drift. When the potential has the value $v_{k}$, the
time interval that one needs to wait till the next event is $\rme^{v_{k}}$. Both
signal $x$ and the potential $v$ change during the jump at time moment $t_{k}$.
One can also consider the case where the time interval between events is random,
with the average equal to $\rme^{v_{k}}$. We can expect that the randomness of the
time interval should not change the PSD of the signal $x_{t}$ at low
frequencies.

In conclusion, we have proposed a pair of coupled nonlinear SDEs
\eref{eq:sde-x-2} and \eref{eq:sde-y-2} that generate the signal $x_{t}$
having the power-law PSD $S(f)\sim f^{-\beta}$ in arbitrarily wide range of
frequencies. The exponent $\beta$ is given by equation \eref{eq:beta-2}. In contrast
to a single nonlinear SDE generating $f^{-\beta}$ noise, the signal $x_{t}$
generated by the proposed pair of SDEs can have almost arbitrary steady-state
PDF. The steady-state PDF of the signal $x_{t}$ is determined only by the
coefficients $a(x)$ and $b(x)$ of the first SDE \eref{eq:sde-x-2}. One can
interpret the first equation \eref{eq:sde-x-2} as describing the fluctuations
of the signal, with the fluctuating rate of change, described by the second
equation \eref{eq:sde-y-2}. Thus, the proposed SDEs exhibit a separation between
the magnitude of the fluctuations of the signal $x_{t}$ and the rate of
fluctuations. We expect that the proposed equations will be useful for the
description of $1/f$ noise in various physical and social systems. In addition,
the equations can be used to numerical generation of $1/f$ noise with the
desired steady-state PDF of the signal.

\section*{References}

\providecommand{\newblock}{}

\end{document}